\documentclass[12pt]{iopart}
\pdfoutput=1

\newcommand{\eqref}[1]{(\ref{#1})}

\usepackage{iopams}
\usepackage{graphicx}
\usepackage{amsfonts}
\usepackage{amsbsy}
\usepackage{amssymb}
\usepackage{color}

\usepackage[top=2cm,bottom=2cm,left=1.5cm,right=1.5cm,columnsep=0.7cm]{geometry}

\usepackage[T1]{fontenc}
\usepackage[utf8]{inputenc}
\usepackage{enumerate}

\usepackage{array}
\usepackage{bm}

\usepackage{subfig}
\usepackage{graphicx}

\usepackage{microtype}

\usepackage{cite}
 
\newcommand\pder[1]{\frac{\partial}{\partial #1}}

\newcommand{\PA}{\tilde{P_s}}
\newcommand{\PV}{P}

\begin{document}

\title[Linear response and correlation of a self-propelled particle in the presence of external fields.]{Linear response and correlation of a self-propelled particle \\in the presence of external fields.}

\author{Lorenzo Caprini$^1$, Umberto Marini Bettolo Marconi $^2$ and Angelo Vulpiani$^{3,4}$}
\address{$^1$ Gran Sasso Science Institute (GSSI), Via. F. Crispi 7, 67100 L'Aquila, Italy}
\address{$^2$ Scuola di Scienze e Tecnologie, Universit\`a di Camerino - via Madonna delle Carceri, 62032, Camerino, Italy}
\address{$^3$ Dipartimento di Fisica, Universit\`a Sapienza - p.le A. Moro 2, 00185, Roma, Italy}
\address{$^4$ Centro Interdisciplinare "B.Segre", Accademia dei Lincei, Roma, Italy}

\ead{ lorenzo.caprini@gssi.it,
angelo.vulpiani@roma1.infn.it}


\begin{abstract} 

We study the non-equilibrium properties of non interacting active Ornstein-Uhlenbeck particles (AOUP) subject to an external nonuniform field using a Fokker-Planck approach with a focus on the linear response and time-correlation functions.
In particular, we compare different methods to compute these functions including the unified colored noise approximation (UCNA).
The AOUP model, described by the position of the particle and the active force acting on it, is usually mapped into a Markovian process, describing the motion of a fictitious
passive particle in terms of its position and velocity, where the effect of the activity is transferred into a position-dependent friction.
We show that the form of the response function of the  AOUP  depends on whether we
put the perturbation on the position and keep unperturbed the active force in the original variables or  perturb the position  and maintain unperturbed the velocity in the transformed variables. Indeed, 
as a result of the change of variables the perturbation on the position becomes a perturbation both on the position and on the fictitious
 velocity.
We test these predictions
by considering the response for three types of convex potentials: quadratic, quartic and double-well potential. Moreover, by  comparing  the response of the AOUP model with the corresponding response of the UCNA model
we conclude that although the stationary properties are fairly well approximated by the UCNA, the non equilibrium properties are not, an effect which is not negligible when the persistence time is large.

\end{abstract}

\maketitle
\section{Introduction}

Self-propelled particles represent a system inherently out of equilibrium as they take energy from the environment, convert it into directed motion
and dissipate it to move in a viscous medium \cite{bechinger2016active,ramaswamy2010mechanics,marchetti2013hydrodynamics}. 
In recent years, a variety of models have been proposed in order to capture both the
stationary and the time-dependent properties of these systems. Among these proposals, we mention the Run and Tumble \cite{Tailleur2009Sedimentation, Nash2010Run, cates2013active}, the active Brownian particle
(ABP) model \cite{solon2015active, romanczuk2012active, Hagen2011} and the Gaussian colored noise model also termed active Ornstein-Uhlenbeck particle (AOUP) model \cite{szamel2014self,flenner2016nonequilibrium,maggi2015multidimensional}. They all describe the overdamped motion of particles subjected to a drag force, due to the solvent, proportional to the particle's velocity, to a deterministic force, $ F$,
 due to an external driving or to interparticle interactions and to the so-called active force or self-propulsion. In the ABP the active force is modeled by a vector of constant norm and whose orientation performs a Brownian
 motion on the unit sphere. The orientation of the self-propulsion has a typical persistence time,$\tau$  i.e.
 it decorrelates with respect to its initial value exponentially as  $\exp(-t/\tau)$.
 The existence of such correlation accounts for the persistence of the trajectories
 which is the distinguishing feature between the standard model of colloidal particles and the one describing self-propelled particles.
 Interestingly, 
in the presence of deterministic forces either due to external fields, such as confining walls or to particle-particle interactions self-propelled particles manifest novel phenomena such as a tendency to cluster \cite{fily2012athermal, buttinoni2013dynamical, bialke2015active} and correlations between the positions and the velocities reflecting their non-equilibrium nature.
 
 The AOUP originates from the necessity of reducing the mathematical complexity of the ABP and is constructed by assuming the same deterministic forces as in the ABP but replacing
 the ABP active force  by an active  force whose components have a Gaussian distribution \cite{fily2012athermal,farage2015effective}. 
The matching between ABP and AOUP is enforced by requiring that 
the active forces of each model  have the same variance and the same exponential time-correlation,
but the AOUP admittedly neglects the non-Gaussian nature of the ABP self-propulsion statistics.
Apart from this non trivial aspect,
the AOUP model has the advantage of lending itself to a simpler analysis and to the possibility of determining the steady state probability distribution function (pdf) of the active particle for small activity \cite{fodor2016far, marconi2017heat}.
Since the AOUP model is formulated as an overdamped particle subject to colored noise, it can be mapped into a new Markovian system, by adding a degree of freedom for each component of the noise. This new enlarged representation allows for the study the problem by using a standard approach based  on the Kramers equation for which several approximate methods
of solution are well-known \cite{Gardiner, risken}.
However, the choice of this enlarged space is not unique since the microstate of a single particle  at a given instant can be identified by its position and velocity or by its position and by the value of the active force acting on it. The two descriptions are
equivalent and for both one can write the corresponding Fokker-Planck equations and the associated approximate steady state 
distribution functions obtained by means of a perturbative analysis in terms of the parameter $\tau^{-1}$.
As far as only the steady state configurational properties are concerned it is possible to derive a closed, non-perturbative expression for the
distribution function by means of the so-called unified colored noise approximation (UCNA) put forward by Hanggi and Jung
\cite{hanggi1995colored,marconi2015towards}.  The static properties of the UCNA have been tested with success in the case
of persistence times not too large, but very little is known about its dynamical properties.
The present study aims to fill this gap by considering the response to a small external perturbation  of a self-propelled particle 
driven by colored noise in the presence of a trapping potential. We shall compare both exact analytic and numerical results obtained by applying the fluctuation dissipation relation (FDR) \cite{marconi2008fluctuation} to the AOUP model for the response 
to an initial displacement of the particle's position with the corresponding quantity computed within the UCNA.
As a byproduct of this study, we obtain and explain a result which contradicts the naive expectation that the positional response
 function should not depend on the choice of the enlarged representation.

The paper is organized as follows. The description of the model and the theoretical results are presented in Section \ref{Theory}. In Sec.~\ref{Numerical Analysis} we report the results of some numerical simulations obtained in the case of anharmonic potentials. Finally, we draw some conclusions in Sec. \ref{Summary and Conclusions}.

\section{Models and theory}
\label{Theory}

We model the effective dynamics for the space coordinates of an assembly of non-interacting active Brownian particles \cite{marconi2015towards,maggi2015multidimensional},  as: 
\begin{eqnarray}
\label{eq:dim_AOUP}
\dot{x}& =& \frac{f(x)}{\gamma}+a, \qquad f(x)=-\frac{d}{dx}\phi(x), \nonumber \\
\dot{a}&=&-\frac{a}{\tau} +\frac{\sqrt{2D}}{\tau}\eta,
\end{eqnarray}
where $x(t)$ is the position of the particle,
$\tau$ is the correlation time, $\gamma$ the drag coefficient, $\phi(x)$ the potential acting on the system
 and the term $a(t)$, also called \emph{active bath}, evolves as an Ornstein-Uhlenbeck process. The stochastic force $\eta(t)$ is a  white noise, i.e. a Gaussian process with zero mean and $\left<\eta(t)\eta(t') \right>=\delta(t-t')$.
The parameter $D$ is the diffusive coefficient due to the activity related 
and fixes the amplitude of  the Ornstein-Uhlenbeck process, via:
\begin{equation}
\left<a(t)a(t') \right> = \frac{D}{\tau} \exp{\left[-\frac{t-t'}{\tau}\right]}.
\end{equation}
In order to proceed further, we will adopt non-dimensional variables for position, velocity, and time
\begin{equation}
X=\frac{x}{l},\qquad V=\frac{v}{v_T},\qquad \bar{t}=t\frac{v_T}{l}
\end{equation}
where 
 $l$ is a suitable length and $v_T=\sqrt{\frac{D}{\tau} } $ is a reference velocity.
 We rescale forces and potential as follows:
\begin{equation}
\qquad F(X) =f(x)\frac{l}{D\gamma},\qquad     U(X)=\frac{\phi(x)}{D\gamma} , \qquad  A=\frac{a}{v_T},
\end{equation}
 and $\zeta =\frac{l}{\tau v_T}$ can be seen as  the ratio 
between the characteristic length of the problem, $l$, such as  the typical length-scale of the external potential $U(x)$, and
the mean square diffusive displacement due to the active bath in a time interval $\tau$.
  Rewriting Eq.\eqref{eq:dim_AOUP}  in terms of these new variables we have:
\begin{eqnarray}
\label{eq:AOUPstoc}
&&\dot{X}=-\frac{U'(X)}{\zeta} + A,  \\
&&\dot{A}=-\zeta A + \sqrt{2\zeta}\xi(\bar t) ,
\end{eqnarray}
where $\langle \xi(\bar t) \xi(\bar t')\rangle=\delta(\bar t- \bar t')$.
In the following, we will use the symbol $t$ for the non-dimensional time.
If the particle is confined to a region of space by a potential $U(X)$,
 $\zeta^{-1}$ represents the ratio between the persistence length 
 and  size of the potential well and the amplitude of the fluctuating force  in reduced units is
 $\lim_{ t \to \infty} \langle A(t) A( t)\rangle= 1$.
For the pdf  ${\tilde P}(X, A, t)$ of the $(X,A)$ variables we have the following Fokker-Planck equation:
\begin{equation}
\pder{t} {\tilde P} - \pder{X}\left(\frac{U'(X)}{\zeta} - A\right) {\tilde P} =\zeta  \frac{\partial}{\partial A}  \left[  \frac{\partial}{\partial A} + A \right] {\tilde P}
\label{fpeactive}
\end{equation} 
whose stationary solution $\PA(X, A)$ is in general unknown a part from simple potentials \cite{risken}.

In order to apply techniques developed for the  Kramers equation it is sometimes convenient 
to use instead of the $(X, A)$ variables  the phase-space variables $(X, V)$ (see for instance
\cite{maggi2015multidimensional,fodor2016far,marconi2017heat}), through the following change of variables:
\begin{eqnarray}
\label{eq:changevariables}
&&V \equiv \dot{X}=-\frac{U'(X)}{\zeta} + A,
\\
 &&X'=X.
\end{eqnarray} 
In this way we recast
the stochastic differential equation \eqref{eq:AOUPstoc} as:
\begin{eqnarray}
\label{eq:kramer}
&& \dot{X}=V\\
&&\dot{V}=  - U'(X) -  \zeta g(X) V+ \sqrt{2T\zeta}\eta
\end{eqnarray}
and the associated Kramers equation for the phase-space distribution $\PV(X, V,  t)$:
\begin{equation}
\label{eq:one_kramers}
\pder{t} \PV + V \pder{X} \PV -U'(X) \pder{V} \PV = \zeta  \pder{V} \left[\pder{V} + g(X)V  \right] P ,
\end{equation}
which means that the activity can be mapped into a space dependent friction term $g(X)=1 + \frac{1}{\zeta^2}\frac{d^2}{dX^2} U(X) $ which depends on $\zeta$. 
The second and third term on the left-hand side represent the streaming terms in the
evolution of the phase-space distribution, whereas the right-hand side describes the dissipative part. Again the stationary distribution $\PV_s(X, V)$ is unknown, in general. Because the Jacobian of the transformation $(X, A)\rightarrow(X', V)$  is unitary, we have:
\begin{equation}
\PV_s(X, V) = \tilde{P}_s(X, A(V, X)).
\end{equation}

We would like to stress that the $X,V$ representation is relevant because it allows us to obtain the distribution function in terms of these variables and to develop an efficient perturbative scheme in powers of the non equilibrium parameter $1/\zeta$.

\subsection{Steady state probability distributions in the extended space}

Among the few cases whose  stationary solution of the Fokker-Plank equation is known, one has the harmonic potential, $U(X)
=\lambda X^2/2$. The steady state distribution 
in the $(X, V)$ variables is a Gaussian, $\PV_s(X,V)\propto e^{-\frac{\beta}{2}\left(\lambda X^2+V^2  \right)}$ with  inverse "effective temperature" $\beta=(1+\lambda/\zeta^2)$, while in the variables $(X, A)$ is the following multivariate Gaussian $\PA(X,A)\propto e^{-\frac{\beta}{2}\left(\lambda X^2+\left(A-\lambda X/\zeta \right)^2 \right)}$.
For a generic potential the stationary pdf in the limit $1/\zeta \ll 1$, has the following approximated
form  (see \cite{fodor2016far,marconi2017heat}):
\begin{eqnarray}
\label{eq:Kdistribution}
&&\PV_s(X, V)\propto e^{-U(X)-\frac{V^2}{2} } \Biggr\{1- \frac{1}{2\zeta^2}\Bigl[U'(X)^2 +V^2U''(X)-3U''(X)\Bigr]
\nonumber\\&&
 + \frac{1}{6\zeta^3} U'''(X) V^3 - \frac{1}{2\zeta^3}U'''(X) V  \Biggr\} +O(\frac{1}{\zeta^4}),
\end{eqnarray}
showing  that positions and velocities are correlated for any finite $\zeta$.
In the $(X, A)$ variables the stationary pdf reads:
\begin{eqnarray}
\label{eq:AOUPdistribution}
&&
\PA(X, A)\propto e^{-U(X)-\frac{\left( A - U'/\zeta \right)^2}{2} } \Biggl\{1- \frac{1}{2\zeta^2}\Bigl[U'(X)^2 +\left(A-\frac{U'(X)}{\zeta}  \right)^2U''(X)-3U''(X)\Bigr]
+\nonumber\\&&
 + \frac{1}{6\zeta^3} U'''(X) \left(A-\frac{U'(X)}{\zeta}\right)^3 - \frac{1}{2\zeta^3}U'''(X) \left(A-\frac{U'(X)}{\zeta}\right)\Biggr\} +O(\frac{1}{\zeta^4})\nonumber\\
\end{eqnarray}
Actually, there are no results in the opposite limit $\zeta \ll 1$, where the persistence time is large.

\subsection{Reduced descriptions: distribution in positional space}
\label{section:Overdampedregimes}
Since in general the analytic solutions of eqs. \eqref{fpeactive} or \eqref{eq:one_kramers}
are not known, it is  common practice to resort to a reduced description involving only the coordinate $X$ for which it is possible to
develop an efficient approximation method. This is the idea behind the reduction of the Kramers equation onto the Smoluchowski
equation and it can be realized by different procedures such as multiple time-scale methods, functional integral techniques or adiabatic procedures.
The unified color noise approximation (UCNA) was developed the first time by Hanggi et Jung by using an adiabatic elimination procedure to study the behavior of particles driven by colored noise~ \cite{hanggi1995colored,h1989colored}  and then recently extended \cite{marconi2015towards, maggi2015multidimensional, marconi2015velocity} for systems of active particles. In the following, we consider two  types of approximations: UCNA and an overdamped limit performed
 directly on the equation \eqref{eq:AOUPstoc}, with the idea of making a comparison among them. We shall study different regimes: $\zeta\gg1$ and $\zeta\ll1$. The first regime corresponds to a small departure from the equilibrium situation determined by the presence of a small $\tau$, while the second regime to the case in which the persistent time is large and is more interesting, because it shows the peculiar features of the active particles, for instance the accumulation of active particles close to confining walls \cite{maggi2015multidimensional}.
In order to gain some insight, we consider the distribution of positions in two special limits corresponding to short
persistence time $\zeta\gg 1$ and to long persistence time $\zeta \ll 1$.

\begin{enumerate}[(i)]
\item $\zeta\gg1$. Let's consider the system of Eq. \eqref{eq:AOUPstoc}; a first approximation consists in neglecting $\dot{A}$. This means that $A$ is well approximated by a white noise and we have:
\begin{equation}
\label{eq:case1_pdf}
\dot{X}  =-\frac{U'(X)}{\zeta} + \sqrt{\frac{2}{\zeta}} \xi, \qquad A= \sqrt{\frac{2}{\zeta}} \xi \qquad \Longrightarrow \qquad P_1(X) \propto e^{-U(X)},
\end{equation}
where $P_1(X)$ is the configurational stationary probability distribution of an equilibrium system.
\item $\zeta \ll1$. In this case, we can neglect $\dot{X}$, therefore the variable $X$ is related to $A$, so that the evolution is given by:
\begin{equation}
\dot{A}=-\zeta A +\sqrt{2\zeta} \xi, \qquad \Longrightarrow \qquad \tilde{P}_2(A) \propto e^{-A^2/2},
\end{equation}
being $\tilde{P}_2(A)$  the stationary pdf for $A$. 
We can obtain a new probability distribution,  $P_2(X)$,  for the variable $X$, since $dA=\frac{U''(X)}{\zeta}dX$
and we have:
\begin{equation}
\label{eq:case2_pdf}
P_2(X) \propto \frac{U''(X)}{\zeta^2}\exp{\left(-\frac{U'(X)^2}{2\zeta^2}\right)} .
\end{equation}
Let us notice that for small $\zeta$,  $P_2(X)$ is a very peaked probability distribution. 
\end{enumerate}
Let us, now, consider the description  based on the variables $(X,V)$ given by  Eq.~\eqref{eq:changevariables} and perform the 
adiabatic elimination of the $V$ variable, i.e. neglecting the acceleration $\dot V$, both for $\zeta \gg 1$ and $\zeta \ll 1$.
We have the following single first order stochastic equation, the well known UCNA equation:
\begin{equation}
\dot{X}=-g(X)^{-1}\frac{U'(X)}{\zeta} + g(X)^{-1}\sqrt{\frac{2}{\zeta}} \xi .
\end{equation}
whose stationary pdf of positions reads:
\begin{equation}
\label{eq:UCNAdistribution}
P_{U}(X) \propto g(X)\exp{\left(-U(X) - \frac{1}{2\zeta^2} U'(X)^2   \right)}.
\end{equation}
Let's remark that
\begin{itemize}
\item  when $\zeta \gg 1$ and $g \sim 1$ we have $P_U \sim P_1$.
\item when $\zeta \ll 1$: since $g \sim U''(X)/\zeta^2$ we have $P_U \sim P_2$ .
\end{itemize}
We note that $P_U$ and $P_1$ are the approximations of the marginal distribution of the system described by the probability distribution given by Eq.~\eqref{eq:Kdistribution}. Indeed, by calling $P_M(X)$ the marginal distribution, with respect to $V$, associated to Eq.~\eqref{eq:Kdistribution}, we have:
\begin{equation}
P_M(X)\equiv\int dV P_s(X, V) = P_U(X) + O\left(\frac{1}{\zeta^4}  \right) = P_1(X) +O\left(\frac{1}{\zeta^2}  \right).
\end{equation}
Therefore, we can say that when $\zeta \gg 1$, the UCNA-model is a better approximation than the model (i)
and there is no reason to use the model (i) instead of the UCNA-model. In particular in the harmonic case the marginal distribution is exactly reproduced by the UCNA.

\subsection{Linear response function}
In this subsection, we shall study the response of our system when we slightly perturb the initial position of the particle
and show that such a procedure yields different results, depending on the variables chosen
 to describe the system. In order to solve this apparent paradox, we first apply the well-known general fluctuation-dissipation relations \cite{falcioni1990correlation,marconi2008fluctuation}, in both representations $(X, V)$ and $(X, A)$. 
 We show that notwithstanding the Jacobian of the transformation is unitary, a perturbation of the position, $X$, in the $A$ representation corresponds to a perturbation involving both the variables ($X$ and $V$) in the $V$ representation.

The response function  $R$ of the AOUP model  was studied by Szamel and Fodor et al. in \cite{fodor2016far,szamel2014self}. In particular Fodor et al. numerically  measured the susceptibility, defined as the time integral of the response and studied the system using the $(X, V)$ coordinates, in the regime of small persistence time. In the present study, instead we directly measure the response of the system, both in the small and in the large persistence time limit. Let's call $R_A(t)$ the mean response of the system that we will compute numerically by adding a
small impulsive force $h(t)=h_0 \delta(t)$, in the first equation of the system \eqref{eq:AOUPstoc}:
\begin{eqnarray}
\label{eq:AOUPstoc_R}
&&\dot{X_h}=-\frac{U'(X_h)}{\zeta} + A_h + h(t), \\
&&\dot{A_h}=-\zeta A_h + \sqrt{2\zeta} \xi,
\end{eqnarray}
where $X_h$ and $A_h$ are the variables of the system in presence of the perturbation. In a similar way we call $R_V(t)$ the response obtained by adding the small force to  equations \eqref{eq:kramer} in the $(X,V)$ representation. The study of $R_A(t)$ and $R_V(t)$ can be numerically performed by computing the following normalized quantity:
\begin{equation} R(t) = \frac{\left<\delta X(t)\right>}{\delta X(0)}=\frac{\left<X_h(t)-X(t)\right>}{X_h(0)- X(0)}.
\end{equation}
Although there are different versions of the FDR for the prediction of $R(t)$ \cite{kubo2012statistical, hanggi1982stochastic}, we focus the attention on the first version of the FDR, independently developed in \cite{A1972} (for a recent work see \cite{Chadauri2012Mod}) and \cite{falcioni1990correlation}.
According to this version of FDR~\cite{marconi2008fluctuation}, we have the response in terms of an average, which involves the stationary pdf:
\begin{equation}
\label{eq:FDR}
R_{V(A)}(t) = -\left<X(t) \left( \pder{X}\log {Prob} \right) \Biggl |_{t=0}  \right>_{V(A)},
\end{equation}
where  depending on the choice of the variables one sets  $Prob$ equal to  $P_s(X, V)$ or  $\tilde P_s (X, A)$  and the average $\left<\cdot\right>$ is performed by using the corresponding stationary pdf
and the symbol $t=0$ means that the function is computed for the variables at $t=0$.

By inserting Eq.~\eqref{eq:Kdistribution} in relation \eqref{eq:FDR},  we obtain the response, $R_V(t)$ in the $(X, V)$ representation:
\begin{eqnarray}
&&
R_V(t)=-\left<X(t) \Biggl(\pder{X}\log{P_s(X, V)} \Biggr)\Biggl |_{t=0}\right>_V =
\nonumber \\&&
\Biggl<X(t)\Biggl(U'(X)+\frac{1}{\zeta^2}\left[ U''(X)U'(X)+\frac{V^2}{2}U'''(X)-\frac{3}{2}U'''(X)\right]
\nonumber \\ &&
-\frac{1}{\zeta^3}\left[\frac{V^3}{6}U''''(X)-\frac{1}{2}VU''''(X) \right] \Biggr)  \Bigg|_{t=0} \Biggr>_V ,
\label{RXV}
\end{eqnarray}
up to the order $O(1/\zeta^4)$, while using Eq.~\eqref{eq:AOUPdistribution} we obtain the response, $R_A(t)$ in the $(X, A)$ representation:
\begin{eqnarray}
&&
R_A(t)=-\left<X(t)\left(\pder{X}\log{\tilde P_s(X, A)}\right)\Bigg|_{t=0}\right>_A =\nonumber \\
&&
\Biggl<X(t) \Biggl(U'(X) - \frac{1}{\zeta} A U''(X)+\frac{1}{\zeta^2}  \left(2 U'(X)U''(X)+\frac{A^2}{2}U'''(X) -\frac{3}{2}U'''(X) \right) +
 \nonumber\\ &&
  \frac{1}{\zeta^3}\left(-A \Bigl[U''(X)^2+U'(X)U'''(X)   \Bigr]   -\frac{A^3}{6}U''''(X)+\frac{A}{2}U''''(X)  \right)   \Biggr)\Bigg|_{t=0}\Biggr>_A,
\label{RXA}
\end{eqnarray}
where $\left< \cdot \right>_A$ denotes the average with respect to $\tilde{P}(X, A)$. By expressing the  response $R_A$ in the variables $(X, V)$ we obtain:
\begin{equation}
\label{eq:response_AOUPXV}
R_A(t)=R_V(t)-\frac{1}{\zeta}\Biggl<X(t) V(0) U''(X(0))\Biggr>_V -\frac{1}{\zeta^3}\left<X(t) V(0) U''(X(0))^2\right>_V \, ,
\end{equation}
showing that  $R_A$ and $R_V$ differ by terms of order $1/\zeta$, which vanish in the limit $1/\zeta \ll 1$. 
Such a result seems somehow counterintuitive:
how  is it possible that the response of the system to an initial perturbation in the $X$ variable  depends on the choice of the coordinates that we use? To explain that,
let's introduce the pdf of the perturbed system $\tilde{P}_s'(X, A)=\tilde{P}_s(X-\delta X_0, A)$ and 
$P_s'(X, V)=P_s(X-\delta X_0, V)$, and let's call $\tilde{W}\left((X_0, A_0)\rightarrow(X, A)\right)$ and $W\left((X_0, V_0)\rightarrow(X, V)\right)$ the transition probabilities from the state at time zero to the one at time $t$ in the coordinates $(X, A)$ and $(X, V)$, respectively. Under the usual hypothesis for the probability distribution  it is easy \cite{marconi2008fluctuation} to show that the response of the position at time $t$, $R_A(t)$, in the variables $(X, A)$, is:
\begin{equation}
<\delta X(t) >_A=\int X \left(\tilde{P}_s'(X_0, A_0) - \tilde{P}_s(X_0, A_0)\right) \tilde{W}\left( (X_0, A_0)\rightarrow (X, A)\right) dX_0 dA_0 dX dA. \nonumber\\
\label{deltax}
\end{equation} 
Since the Jacobian of the transformation is unitary , we can switch from the variables $(X, A)$ to  $(X, V)$:
\begin{eqnarray}
&&<\delta X(t) >_A= \nonumber\\
&&\int X \Bigl[P_s'(X_0, V_0(X_0 ,A_0)) - P_s(X_0, V_0(X_0, A_0))\Bigr] W\left( (X_0, V_0)\to (X, V)\right) dX_0 dV_0 dX dV.
\end{eqnarray} 
For a small increment $\delta X_0$ we have:
\begin{eqnarray}
&&
P_s'(X, V(X ,A)) - P_s(X, V(X, A)) = -\delta X_0\left[ \frac{\partial}{\partial X}P_s(X,V(X ,A))+\frac{\partial V(X,A)}{\partial X}\frac{\partial P_s}{\partial V}   \right]
\nonumber\\&&
=-\delta X_0 \frac{d}{dX}P_s(X, V(X, A)).
\end{eqnarray}
This version of the FDR involves a total derivative, which acts also on the velocity. This means that 
the response $R_A(t)=\frac{<\delta X(t) >_A}{\delta X_0}$ is given by:
\begin{eqnarray}
\label{eq:GeneralFDR}
&&
R_A(t)=-\left<X(t)\left(\frac{d}{dX}\log{P_s(X, V(A, X)})\right)\Bigg|_{t=0}\right>_V 
\nonumber\\&&
\neq-\left<X(t)\left(\pder{X}\log{P_s(X, V(A, X)})\right)\Bigg|_{t=0}\right>_V=R_V(t),
\end{eqnarray}
where the averages are performed by using $P_s(X, V)$.  
In other words we can say that, since $A$ depends on $X$ and $V$, the perturbation $(X, V) \rightarrow (X+\delta X_0, V)$ is not equivalent to the perturbation $(X, A) \rightarrow (X+\delta X_0, A)$.

\section{Results}
\label{Numerical Analysis}
In the following, we shall present some results illustrating the predictions of the theory  
in some simple cases. We have performed the simulations for three different potentials $U(X)$: (a) harmonic potential $U(X) = \lambda X^2/2$, (b) quartic potential $U(X) = \lambda X^4/4$,  (c) double well potential $U(X) =\lambda ( X^4/2-X^2/2)$.

The numerical computations of $R(t)$, both from data and FDR Eq.\eqref{eq:FDR}, were performed using the Euler-Maruyama method \cite{ToralColen_book}, neglecting order $(\Delta t)^{5/2}$.

\subsection{Response in the limit $\zeta \ll 1$}

In the case (a) the probability distribution of the system can be computed exactly and therefore we have an exact expression for the response: $R_A(t)\sim e^{-t \lambda/\zeta}$.
In the cases (b) and (c)
we  know the probability distribution as a series in powers of $1/\zeta\ll 1$ so that we can obtain the FDR only perturbatively. Therefore, the numerical approach is necessary when the limit $\zeta\gg 1$ doesn't hold.  In Fig.~\ref{fig:threepotentials} we show $R_A(t)$  for the three different potentials and different values of $\zeta$. Let us first discuss the case (a) and (b): 
\begin{figure}[!h]
\centering
\subfloat[]{\includegraphics[width=0.33\linewidth,keepaspectratio]{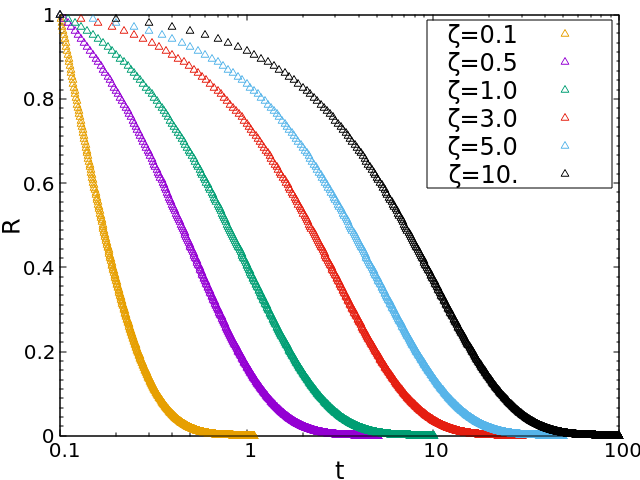}\label{fig:1KvsA}}
\subfloat[]{\includegraphics[width=0.33\linewidth,keepaspectratio]{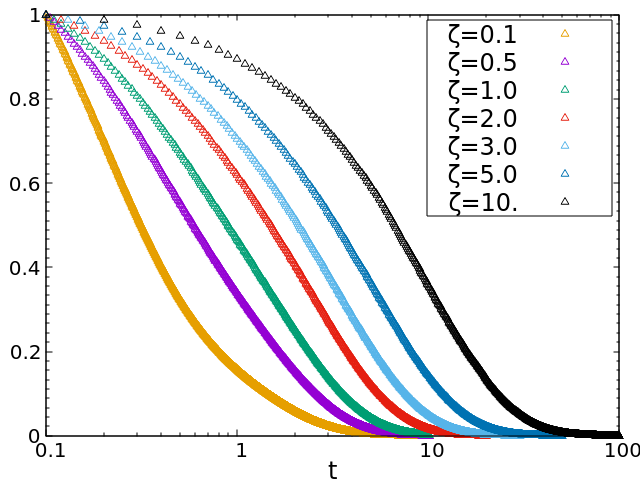}\label{fig:3KvsA}}
\subfloat[]{\includegraphics[width=0.33\linewidth,keepaspectratio]{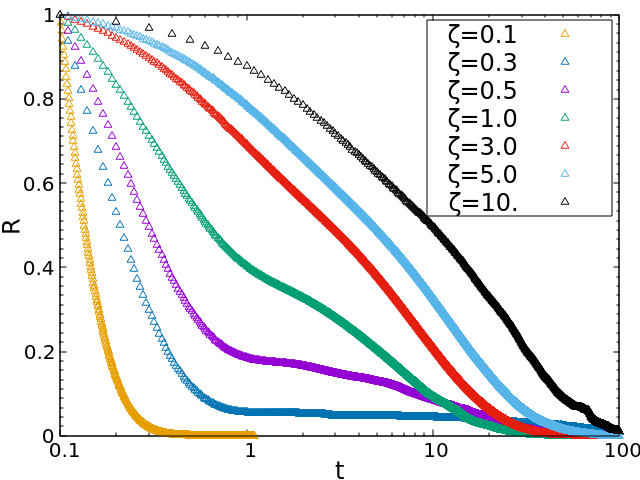}\label{fig:5KvsA}}
\caption{Responses $R_A(t)$ for the AOUP model computed, via numerical simulation, for different values of $\zeta$ 
by setting $\lambda=1$. 
In figure (a) are plotted the responses for the harmonic potential $\lambda X^2/2$. The plots (b) and (c) correspond to a quartic potential $\lambda X^4/4$ and a double well potential $\lambda (X^4/4 - X^2/2)$, respectively.} 
\label{fig:threepotentials}
\end{figure}
when $\zeta$ is large, we are near the delta correlated noise, closed to the equilibrium situation. Therefore the shape of the potential does not change the form of the response, which decays roughly as an exponential.
When $\zeta\sim 1$ or $\zeta\ll 1$ the results relative to the two potentials display large differences
increasing as $\zeta$ decreases.
In particular, when the attractive force becomes stronger, the response becomes slower, as we can see in Fig.~\ref{fig:threepotentials}. 
This is a consequence of the departure of the system from thermodynamic equilibrium: indeed the detailed balance holds only in the harmonic case \cite{marconi2017heat} when the drag coefficient in the $(X, V)$ variables is constant. Otherwise, $g(X)$ is not constant,  and decreasing $\zeta$ the system goes far from equilibrium. 
Indeed, where $\zeta$ is small, in the harmonic case the shape obtained is the one predicted by the theoretical computations $\sim e^{-\lambda t/\zeta}$.  In the other cases we can distinguish between two regimes: up to $t\sim 1$ (in 
dimensional units this corresponds to $t\sim l/v_T$) there is \emph{fast} relaxation, while for $t\gg1$ there is a relaxation with an effective characteristic time $t_{slow} \gg \zeta$. 
The presence of these two regimes is a non-equilibrium effect and is more evident when the activity is large ($\zeta$ small). 
The presence of two times scales, when the system is far from equilibrium, is a clear consequence of the accumulation of particles near the confining walls \cite{maggi2015multidimensional}.  This means that, even if the potential applied has a single well, the particle in the steady state experiences an effective double well potential. 
Such an observation is confirmed by the shape of the stationary pdf  $P_U(X)$ in the UCNA-approximation.
 Phenomenologically, the drift term takes different values depending on the position of the particle: when $X$ is near the minimum of the potential the effective drift force  is proportional to $\sim \zeta V\ll V$, being $U''(X)/\zeta \sim 0$, and the particle moves just because of the deterministic force. For $X$ far from the minimum, the drift force is proportional to $\sim  V U''(X)/\zeta \gg V$, which means that the particle experiences a big Stokes force and moves very slowly. For $X$ far from the minimum, the deterministic force is very big and steadily pushes the particle towards the minimum, preventing the particle from going too far. The balance between these two effects leads to a situation where the most probable value of the position does not coincide with the minima of the potential. This fact explains why the decay of the response function displays two different time-regimes, even in the presence of a single well potential $U(X)\propto X^{2n}$ with $n>1$ .
 Let us remark that this mechanism acts only when the detailed balance does not hold, as in the case where the curvature
 $U''(X)$ is not constant \cite{marconi2017heat}.

Finally, we consider the the double well potential (case (c)). In order to gain some insight, let us consider the situation where the noise is delta correlated: the response function displays two different decay behaviors (roughly exponential), in the first stage the typical decay time is associated with the relaxation in one of the two wells and is determined by 
the curvature of the potential, $U^{''}(X_{min})$. For longer times the jumps of the particle between the two minima are relevant and the mean first passage time is determined by Kramers' formula \cite{Gardiner}. If the persistence time, $\tau$, is not very  small is not easy to extend  the above argument, however,  in Fig.~\ref{fig:threepotentials} (c) which displays the behavior of the response function versus $t$, it is quite evident the presence of two different characteristic time scales.
When the persistence time becomes larger the second relaxation becomes slower as clearly indicated by 
the plot of Fig.~\ref{fig:threepotentials} (c), as if the effective barrier becomes higher. 

\subsection{The UCNA response function}
It is known that the UCNA model well describes 
all the stationary properties of the system  both for $1/\zeta \ll1$ and $\zeta \ll1$.
This state of affairs is no longer true  for the time-dependent dynamical properties such as the response to a small perturbation. 

By using the FDR for a system under the action of a generic potential $U(X)$, we easily obtain the following expressions for the responses in the three cases, denoted by a subscript.
\begin{enumerate}
\item if $\zeta \gg 1$ from eq. \eqref{RXV} we have:
\begin{equation}
R_1 = \left<X(t) U'(X(0))  \right>_V.
\end{equation}
\item while for $\zeta \ll 1$ the response is 
\begin{equation}
R_2(t) =\frac{1}{\zeta^2}\left<X(t) U'(X(0))U''(X(0))  \right>_V - \left<X(t) \frac{U'''(X(0))}{U''(X(0))}\right>_V.
\end{equation}
\item 
and within the  UCNA we have:
\begin{equation}
R_U(t) = -\left<X(t)\left(\pder{X}\log{P_U(X)}\right)\Bigg|_{X=X(0)}  \right>_U,
\label{eq:RU}
\end{equation}
and explicitly:
\begin{equation}
R_U(t) = \frac{1}{\zeta^2}\left<X(t) U'(X(0))U''(X(0))   \right>_U +  \left<X(t) U'(X(0))   \right>_U - \left<X(t) \frac{U'''(X(0))}{\zeta^2+U''(X(0))}\right>_U.
\end{equation}
\end{enumerate}
where the subscript $U$ means that the average is with respect to the UCNA steady state distribution, $P_U(X)$
given by \eqref{eq:UCNAdistribution}.

\subsection{Response function in the presence of a quadratic potential}
In the harmonic case, $U(X)=\lambda X^2/2$, we can apply the FDR for all values of $\zeta$ without approximations
and obtain from Eq.~\eqref{RXV}:
\begin{equation}
R_V(t) = \beta \lambda \left<X(t) X(0)  \right>_V
\label{eq:xx}
\end{equation}
and from Eq.~\eqref{eq:response_AOUPXV}:
\begin{equation}
R_A(t)=\beta \lambda \Bigl(\left<X(t) X(0)  \right>_V-  \frac{1}{\zeta}\left<X(t)V(0) \right>_V \Bigr) = R_V(t) -\beta  \frac{\lambda}{\zeta}\left<X(t)V(0) \right>_V .
\label{eq:xv}
\end{equation}
In general, the two responses are not the same, except in the limit $\zeta \rightarrow \infty$ which corresponds to the $\delta$-correlated case. 
As shown in the Appendix the correlation functions appearing in r.h.s. of Eqs. \eqref{eq:xx} and \eqref{eq:xv} are given by:
\begin{equation}
\label{eq:XcorrelationF}
\beta\left<X(t)X(0)\right>_V = \frac{1}{\zeta -\lambda/\zeta} \left[\frac{\zeta}{\lambda} e^{-\lambda t/\zeta} - \frac{1}{\zeta}e^{-t\zeta}  \right],\\
\end{equation}
and
\begin{equation}
\label{eq:VcorrelationF}
\beta\left<X(t)V(0)\right>_V =  \frac{1}{\zeta -\lambda/\zeta} \left[ e^{-\lambda t/\zeta} - e^{-t\zeta}  \right].
\end{equation}
Finally, the response functions read:
\begin{equation}
\label{eq:V_response}
R_V(t) = \frac{\lambda}{\zeta-\frac{\lambda}{\zeta}} \left( \frac{\zeta}{\lambda} e^{-\lambda t/\zeta} - \frac{1}{\zeta}e^{-t \zeta} \right)\\
\end{equation}
and
\begin{equation}
\label{eq:exact_response}
R_A(t) = e^{-\lambda t/\zeta},\\
\end{equation}
Perhaps contrary to  intuition, the two responses are different for $\zeta$ not too large. For large $\zeta$ the two responses are very close.  This is an effect of the memory: indeed, small $\zeta$ means big correlation time, being $\zeta\sim 1/\tau$.

\begin{figure}[!h]
\centering
\subfloat[]{\includegraphics[width=0.35\linewidth,keepaspectratio]{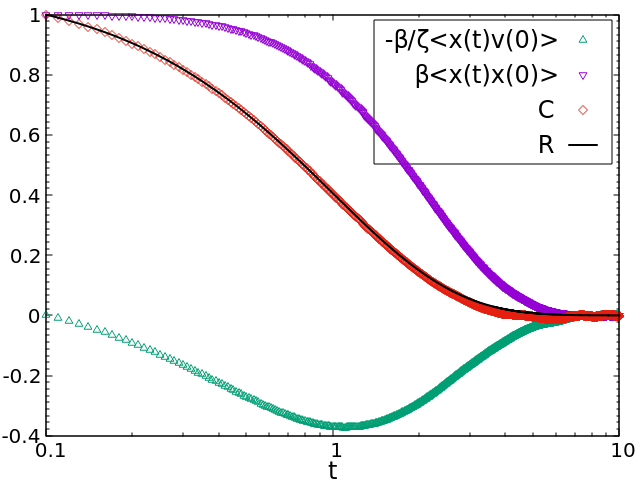}\label{fig:1KA}}
\subfloat[]{\includegraphics[width=0.35\linewidth,keepaspectratio]{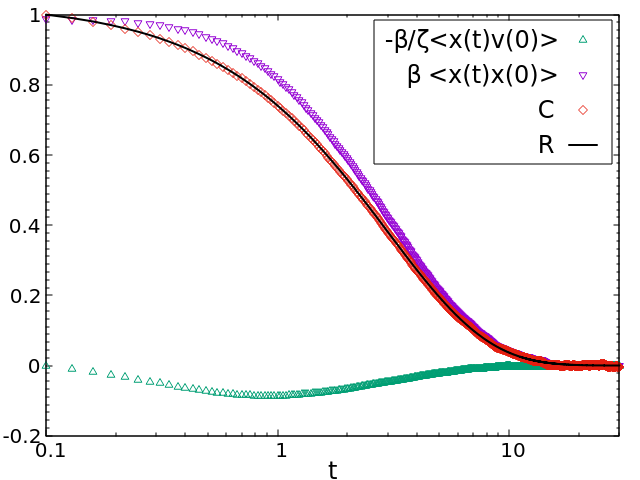}\label{fig:3KA}}
\caption{
Response function $R_A(t)$ computed via numerical simulation for a harmonic potential $\lambda X^2/2$ with $\lambda=1$ (black line). The red diamonds (C) represent the sum of the correlation functions  given by the Eq.~\eqref{eq:xv}, i.e.
$R_A(t)=\beta \lambda \Bigl(\left<X(t) X(0)  \right>_V-  \frac{1}{\zeta}\left<X(t)V(0) \right>_V \Bigr)$, a test employed 
in order to verify FDR.
The green triangles represent the correlation $\beta\left<X(t)V(0)\right>_V$ (Eq.~\eqref{eq:VcorrelationF}) and
the violet inverse triangles represent the correlation $\beta\left<X(t)X(0)\right>_V$ (Eq.~\eqref{eq:XcorrelationF}) and . 
Panel (a) corresponds to $\zeta=1$ and panel (b) to $\zeta=3$.}
\label{fig:KR}
\end{figure}

Consider now the response function as predicted by the UCNA theory in the harmonic case.
By using the FDR given by Eq. \eqref{eq:RU} together with the UCNA stationary probability distribution
 \eqref{eq:UCNAdistribution} and the correlation function 
$$
<X(t)X(0)>_U=\frac{1}{\beta\lambda} \, \exp{\left(-\frac{\lambda t}{\zeta + \lambda/\zeta} \right)} ,
$$ the response function turns out to be:
\begin{equation}
R_U(t)  =  \exp{\left(-\frac{\lambda t}{\zeta + \lambda/\zeta} \right)}.
\end{equation}
Let's observe that this response is invariant for $\zeta \rightarrow \lambda/\zeta$. Then:
\begin{itemize}
\item $\zeta\gg1\quad \Longrightarrow \quad R_U(t) \sim \exp{\left(-\frac{\lambda t}{\zeta}\right)}$, which is consistent with the  response $R_A(t)$, given by  Eq~\eqref{eq:exact_response}.
\item $\zeta\ll1\quad \Longrightarrow \quad R_U(t) \sim \exp{\left(-t\,\zeta\right)}$, which is not correct. 
Smaller $\zeta$ means a slower response, in disagreement with the result for $R_A(t)$, given by the Eq.~\eqref{eq:exact_response}.
\end{itemize}

\subsection{Response function with varying $\zeta$}
\begin{figure}[!h]
\centering
\subfloat[]{\includegraphics[width=0.45\linewidth,keepaspectratio]{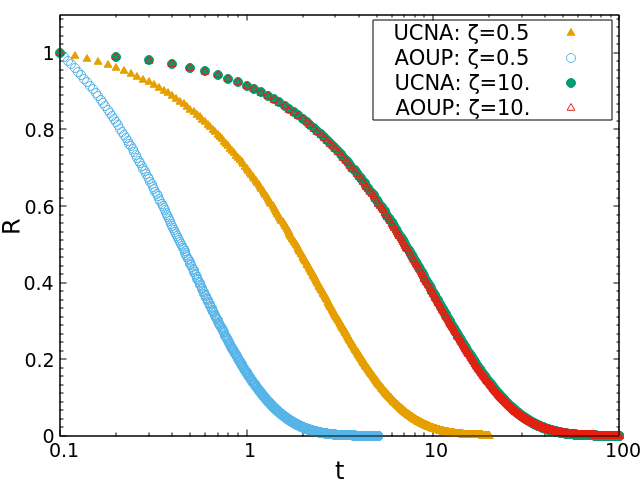}\label{fig:AOUPUCNA_10e05}}
\subfloat[]{\includegraphics[width=0.45\linewidth,keepaspectratio]{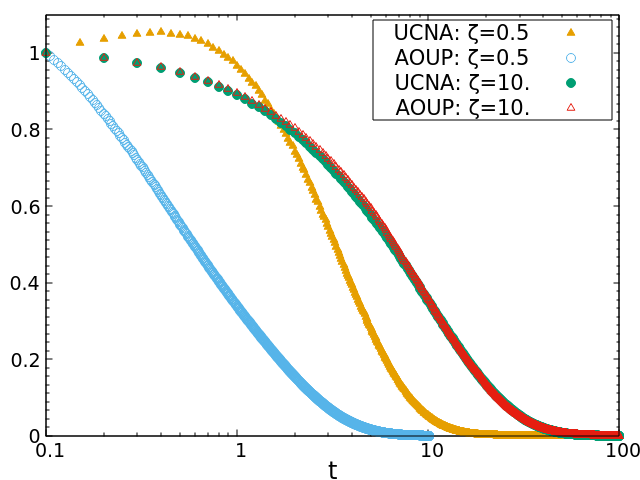}\label{fig:four_AOUPUCNA_05e10}}
\caption{Responses functions computed via numerical simulations from the AOUP ($R_A(t)$) model and the UCNA-model
($R_U(t)$), for different values of $\zeta$: $\zeta=10, 0.5$ and $\lambda=1$. The graphs are obtained for systems under the action of a harmonic potential $U=\lambda X^2/2$ ( panel (a)) and a potential $U=\lambda X^4/4$ (panel (b)).}\label{fig:10e05}
\end{figure}

Let us show the responses, numerically computed, for harmonic and quartic potentials,  $U(X)=\lambda X^2/2$  and  $U(X)=\lambda X^4/4$, respectively.
In the harmonic case, the simulations are only intended as a check of the numerical codes. In the quartic case, we have only a perturbative result in power of $1/\zeta \ll1$ for the probability distribution function (see eq. \eqref{eq:Kdistribution}). In general,  it is difficult to predict the 
response in the small-$\zeta$ regime and we need a numerical study.
In fig.  \ref{fig:10e05} we show a comparison between the response of the AOUP-model and the UCNA. For $\zeta \gg 1$ the UCNA is a good approximation of the AOUP both in the case of a harmonic potential (Fig. \ref{fig:AOUPUCNA_10e05}) and of a quartic potential (Fig. \ref{fig:four_AOUPUCNA_05e10}). When $\zeta$ becomes smaller, the situation is completely different: in the harmonic case, as predicted by the theoretical computation, the response $R_A(t)$ becomes slower, according to the invariance $\zeta \rightarrow \lambda/\zeta$. Even in this simple case, the UCNA is not able to reproduce the response of the AOUP system for $\zeta \ll 1$. The scenario is similar in the case of the quartic potential $U(X) = \lambda X^4/4$. 
\begin{figure}[!h]
\centering
\subfloat[]{\includegraphics[width=0.33\linewidth,keepaspectratio]{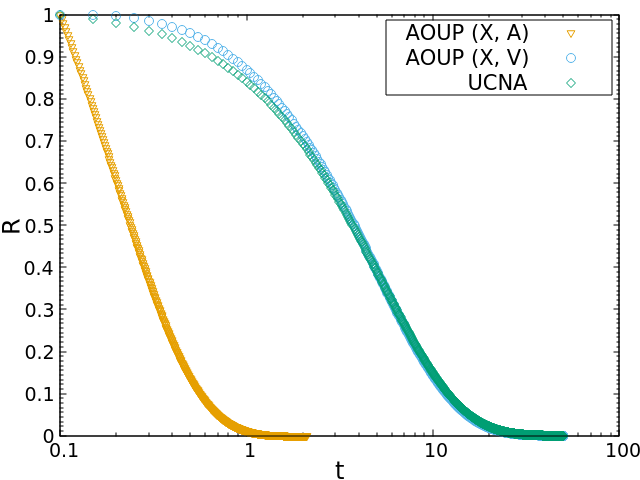}\label{fig:two_02AOUPkramerUCNA.png}}
\subfloat[]{\includegraphics[width=0.33\linewidth,keepaspectratio]{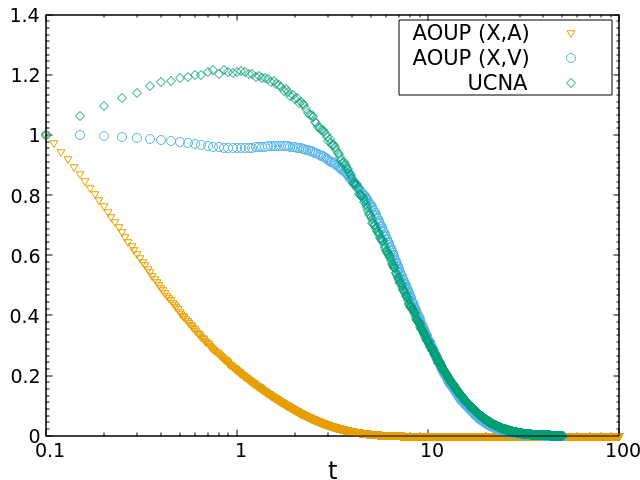}\label{fig:four_02AOUPkramerUCNA.png}}
\subfloat[]{\includegraphics[width=0.33\linewidth,keepaspectratio]{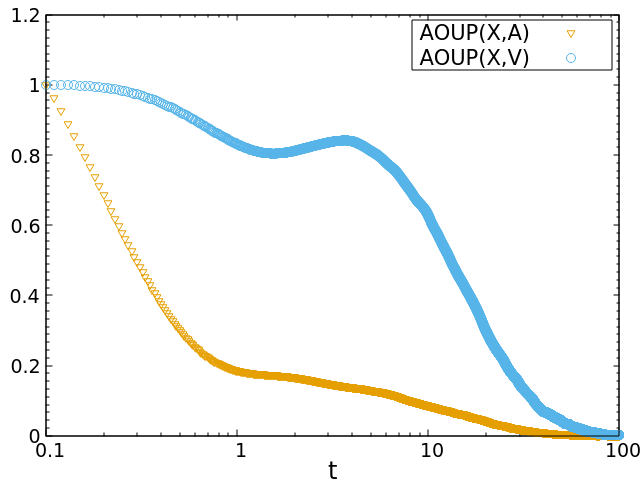}\label{fig:05kramerAOUP.png}}
\caption{Response functions $R_A(t)$ and $R_V(t)$  computed via numerical simulations from the AOUP model
and response $R_U(t)$ of the UCNA in the case of  small $\zeta$: $\zeta=0.2$ and $\lambda=1$. The graphs (a)  and (b) are obtained for systems under the action of a harmonic potential $U=\lambda X^2/2$ and a quartic one $U=\lambda X^4/4$, respectively. The graph (c) is obtained for the double well potential $U=\lambda (X^4/4- X^2/2)$, for $\zeta=0.5$. }\label{fig:02AOUPkramerUCNA}
\end{figure}
Moreover, we observe that the response $R_U(t)$ computed within the UCNA can be seen only as an approximation of the response $R_V(t)$ computed from Eq.\eqref{RXV}.

In Fig. \ref{fig:02AOUPkramerUCNA} we show a comparison between the response functions $R_V$, $R_A$ and $R_U$. 
We have that the responses $R_V(t)$ and $R_A(t)$ display marked differences when $\zeta/\lambda \ll 1$ and $R_A(t)$ decays much faster. 
The explanation in the harmonic case comes from the FDR Eq.\eqref{eq:xv}: Indeed the correlation between $X$ and $V$ plays an important role only for $\zeta$ small enough, giving a non-vanishing negative contribution. This means that the coupling between the $X$ and $V$ in the Eq.\eqref{eq:xv}  is responsible for the faster decay of $R_A(t)$ with respect to $R_V(t)$, as is shown in Fig.\ref{fig:KR}.
On the other hand, $R_U(t)$  in the harmonic case is very close to $R_V(t)$, but in the initial stage
the UCNA response is non-monotonic and only in a later stage behaves very closely to $R_V(t)$.
In fact, $R_U(t)$ is only an approximation of $R_V(t)$ and fails in the limit $\zeta \ll 1$. This has been directly verified both for the harmonic potential (Fig. \ref{fig:two_02AOUPkramerUCNA.png}) and the quartic potential (Fig. \ref{fig:four_02AOUPkramerUCNA.png}). 
In Fig.\ref{fig:02AOUPkramerUCNA} we compare $R_V$ and $R_A$ in the case of a double well potential when $\zeta$ is small and so the system is far from equilibrium. As expected by the previous cases, $R_V(t)$ is slower than $R_A(t)$. Both responses show a first exponential relaxation for $t\sim 1$ and a relaxation slower than an exponential for a much longer time.

\section{Summary and Conclusions}
\label{Summary and Conclusions}
In this paper, we studied the response of a one-dimensional system of non-interacting AOUP under the action of an external potential. We have
shown that, at variance with the equilibrium case which applies to passive particles, the standard formula connecting the response function after an initial perturbation in the particle position, $X$, to the  partial derivative of the stationary 
phase-space distribution function, $P(X,V)$, with respect to  $X$ has to be modified in the case of active particles.
Such a modification is necessary due to the dependence of the velocity of the particle, $V$, on the position $X$, a distinguishing feature of the
active dynamics.
The relevance of the derived formula  is important when the persistence time $\tau$ is large.
 In order to validate our claims, we studied the analytically solvable case of a quadratic potential and by numerical methods the case of non quadratic potentials
and compared  the response of the AOUP system with the response in the overdamped regime corresponding to the UCNA  both for small and large persistence time.
This analysis shows that although the stationary properties are well approximated by the UCNA in both cases, this is not true regarding the dynamical properties. In particular, in the case of the response, the UCNA is a good approximation only when the persistence time is small. 
Finally, the present study has shown that when the persistence time is large enough, even in the case of a single well $U(X)\propto X^{2n}$ with $n>1$, the response function  relaxation is characterized by two time scales. This result is a clear manifestation of the
non equilibrium nature of the system and appears only when the detailed balance does not hold.

\ack

We thank A. Puglisi for useful discussions and the anonymous referees for constructive criticisms.

\section{Appendix: Exact computation of correlations and response functions for the harmonic potential}\label{Appendix:HarmonicOscillator}

In the case of the harmonic potential the exact stationary  probability distribution, $P_s(X,V)$, is well known, being a Gaussian 
with respect to both variables so that we can compute the time-dependent correlations and the responses $R_{A(V)}$
using the FDR. 
Indeed, by first multiplying by $X(0)$  the system of equations \eqref{eq:AOUPstoc} 
and then taking their average with respect to the steady distribution $P_s(X,V)$ , we  obtain a system of ordinary differential equation for the correlation functions.
Let's start from the evolution equations for the averages
\begin{eqnarray}
&&\frac{d}{dt}\left<X(t)X(0)\right>_V = \left<A(t)X(0)\right>_V -  \frac{\lambda}{\zeta}\left<X(t)X(0)\right>_V, \\
&&\frac{d}{dt}\left<A(t)X(0)\right>_V =-\zeta\left<A(t) X(0) \right>_V .
\end{eqnarray}
We can solve 
\begin{equation}
\left<A(t)X(0)\right>_V =  B e^{-\zeta t},
\end{equation}
where $B$ is a constant to be fixed by the initial conditions. By substituting we get:
\begin{equation}
\frac{d}{dt}\left<X(t)X(0)\right>_V =  -  \frac{\lambda}{\zeta}\left<X(t)X(0)\right>_V + B e^{-\zeta t},
\end{equation}
whose solution is:
\begin{equation}
\left<X(t)X(0)\right>_V = C e^{-\lambda t/\zeta} - \frac{B}{\zeta-\frac{\lambda}{\zeta}}e^{-\zeta t},
\end{equation}
where $C$ is a second costant to be determined. Let us choose the following steady state initial conditions:
\begin{eqnarray}
&&\lambda \left<X(0)X(0)\right>_V= \frac{1}{\beta },\\
&&\left<V(0)X(0)\right>_V =  \left(\left<A(0)X(0)\right>_V - \frac{\lambda}{\zeta}\left<X(0)X(0)\right>_V  \right) = 0,
\end{eqnarray}
which mean that initially the "potential" energy $\lambda X^2/2$ obeys an equipartition principle and
 the velocity is not correlated with the position. In this way we can easily determine the constants: $B=1/(\zeta \beta)$ and $\beta C=(\zeta/\lambda)/(\zeta-\lambda/\zeta)$. Finally, we have:
\begin{equation}
\beta\left<X(t)X(0)\right>_V = \frac{1}{\zeta -\lambda/\zeta} \left[\frac{\zeta}{\lambda} e^{-\lambda t/\zeta} - \frac{1}{\zeta}e^{-\zeta t}  \right]=\frac{1}{\lambda} R_V(t).
\end{equation}
We can, now, easily compute $\left<V(t)X(0)\right>$:
\begin{equation}
\beta\left<V(t)X(0)\right>_V = \beta\frac{d}{dt}\left<X(t)X(0)\right>_V = -  \frac{1}{\zeta -\lambda/\zeta} \left[ e^{-\lambda t/\zeta} - e^{-\zeta t}  \right].
\end{equation}
By using the reversibility condition $\left<V(t)X(0)\right>_V=-\left<X(t)V(0)\right>_V$, the response of the AOUP system reads:
\begin{equation}
R_A(t)= \beta\lambda \left<X(t)X(0)\right>_V - \beta\frac{\lambda}{\zeta}\left<X(t)V(0)\right>_V = e^{-\lambda t/ \zeta},
\end{equation}
which is our exact result.

 By the same methods we can compute  the correlation functions $\left<A(t)X(0)\right>_V$ and $\left<X(t)A(0)\right>_V$.
 Since the harmonic oscillator driven by colored noise obeys the detailed balance condition,  if the variable $A(t)$ had a 
 a well defined parity under time-reversal  one would obtain the relation $$\left<A(t)X(0)\right>_V =  \pm \left<X(t)A(0)\right>_V .$$ 
Indeed, this is not the case and as a matter of fact the result is:
\begin{eqnarray}
&&\left<A(t)X(0)\right>_V=\left<V(t)X(0)\right>_V +\frac{\lambda}{\zeta}\left<X(t)X(0)\right>_V=\frac{1}{\beta}\frac{e^{-t\zeta}}{\zeta}, \nonumber\\
&&\left<X(t)A(0)\right>_V
=\left<X(t)V(0)\right>_V +\frac{\lambda}{\zeta}\left<X(t)X(0)\right>_V
=\frac{1}{\beta} \frac{1}{\zeta-\lambda/\zeta} \left( 2\,e^{-\lambda t/\zeta } - (1+\lambda/\zeta^2) e^{-\zeta t}\right). \nonumber
\end{eqnarray}


\end{document}